
\input harvmac

\def \dvps {{\dot \varphi}^2}
\def \dls {{\dot \lambda }^2}

\def \const {\rm const}

\def \dlis {{{\dot \l}_i}^2}

\def \dop  {\dot {\phi} }
\def \ddp {\ddot \phi }

\def \eq#1 {\eqno{(#1)}}
\def \e {\rm e}

\def \evp {\rm e^\varphi }

\def \e#1 {{\rm e}^{#1}}

\def \a {\alpha}
\def \b {\beta}

\def \ln {\rm ln }

\def \l {\lambda}
\def \p {\phi}
\def \vp {\varphi}

\def \r {\rho}

\def \s {\sigma}

\def \b {\beta}

\def \pa {\partial}

\def \sqG {\sqrt {-G}}
\def \Goo {G_{00}}

\def \dl {\dot \lambda}
\def \ddl {\ddot \lambda}
\def \dvp {\dot \varphi}
\def \ddvp {\ddot \varphi}
\def \sm {\sum_{i=1}^{N}}

\def\NP {  Nucl. Phys.\ }
\def \PL { Phys. Lett.\ }
\def \MPL { Mod. Phys. Lett.\ }
\def \PRL {  Phys. Rev. Lett.\ }
\def \PR  { Phys. Rev. \ }

\Title{\vbox{\baselineskip12pt\hbox{HUTP-91/A049}\hbox{JHU-TIPAC-910028}}}
{\vbox{\centerline{Elements of String Cosmology}}}

\centerline{ A. A. Tseytlin\footnote{$^\dagger$}
{On leave of absence from the Department of Theoretical Physics, P. N.
Lebedev Physics Institute, Moscow 117924, USSR.
Address from  October 1, 1991 : DAMTP, University of Cambridge, Cambridge
CB3 9EW   and Trinity College, Cambridge, CB2 1TQ, Cambridge, United
Kingdom.}}
\vskip .1 in
\centerline{\it Department of Physics and Astronomy}
\centerline{\it The Johns Hopkins University}
\centerline{\it Baltimore, Maryland 21218, USA}
\baselineskip=20pt plus 2pt minus 2pt
\vskip .1in
\centerline{C. Vafa}
\centerline{\it Lyman Laboratory of Physics}
\centerline{\it Harvard University}
\centerline{\it Cambrdige, MA 02138, USA}
Aspects of string cosmology for critical and non-critical
strings are discussed emphasizing the necessity to account for
  the dilaton dynamics for a proper incorporation of ``large - small"
duality.  This
drastically modifies the intuition one has with Einstein's gravity.
For example winding modes, even though contribute to energy density, oppose
expansion and if not annihilated will stop the expansion.
Moreover we find that
the radiation dominated era of the standard cosmology emerges
quite naturally in string cosmology.
Our analysis of non-critical string cosmology
provides a reinterpretation of the (universal cover of the)
recently studied two dimensional
black hole solution as a conformal realization of cosmological
solutions found previously by Mueller.

\Date{9/91} 


\newsec{ Introduction}
Despite tremendous progress in our understanding of fundamental strings
in the past decade, we are still very far from a single quantitative
prediction to be observed in experiments!  The main reason for this
unsatisfactory state of affairs is that the natural scale
in which the stringy effects become important (Planck scale) is
much smaller than the scale we can probe in high energy scattering
experiments.  However, it has been known for a while that
cosmology provides a window to test fundamental physics,
since the scales of particle physics theories become relevant
 as the universe evolves in its early history.
It is our belief that the most likely area for a confrontation
between string theory and experiments is through extracting cosmological
consequences of string theory which may be measurable today.  Many
of the riddles of astrophysics, for example `dark matter' and large
scale structure of the universe may turn out to be related to fundamental
strings.  It is with such a program in mind that we study
string cosmology in this paper.

To see the effects of strings
we have to trace back the universe to its earliest times.  Here
already string theory provides us with a surprise: strings in very
small spaces behave the same way as strings in very large spaces [1,2].
This is true at least in the simplest known cases (toroidal backgrounds)
and is suspected to be a  general principle of string theory,
because strings should be `blind' to scales smaller than their own
natural scale.  Thus the universe
does not start at a zero size, as that would be equivalent
to infinitely large universe, rather we should start with a universe
of the size of the string scale, i.e., the Planck length.  This $a\rightarrow
1/a$ duality (in Planck units)
 already hints of a major modification to Einstein equations
at early times, as ordinary theory of general relativity is not
invariant under such a transformation.  It has been suggested that, roughly
speaking, string theory necessitates the existence of two kinds of matter
 fields, ``dual'' to each other.  For large radii, one type becomes
massless and dominates, while for small radii the other type becomes massless
and dominates.  This suggests an effective ``doubling" of
the number of space dimensions (see e.g. [2,3]).  Needless to say to have
an exact treatment of {\it dynamics} in
such a situation is beyond our scope at the present
time.  This was one of the main reasons why the cosmological ideas
in [2] were rather difficult to test, as the
dynamical equations governing string cosmology were not properly understood.

It turns out that it is possible to introduce a
consistent dynamics which describes slowly evolving states of strings
in a duality invariant manner.
This involves, in particular,  taking
into account the dilaton field which transforms under the duality [4-7].

It is a fundamental consequence of string theory that the gravitational
interaction is effectively described not by the metric alone but by the
coupled system of the metric and the dilaton. Dilaton of course should get
a mass to avoid a scalar component of gravitational interaction of massive
bodies but this probably happens at a later
stage of evolution of the Universe. That implies that one
 cannot not ignore the dilaton dynamics at early ``stringy" phases of
cosmological evolution. As we shall see, the
 assumption that dilaton is constant
is in general inconsistent with the cosmological equations
describing evolution of matter with the ``stringy" equation of state\foot{
 Dilaton dynamics was ignored in most previous discussions of string
cosmology at finite temperature
 (see e.g. [8] and refs. there). Dilaton was not set equal to constant in
ref.[9] where, however, the space was not taken to be compact and hence
many of the stringy effects, and in particular duality invariance
was not discussed.}.

The main body of this paper is devoted to the discussion of the duality
invariant cosmological
equations describing the
evolution of the ``scales" of the metric and  the dilaton.
 We shall study their solutions emphasizing departures
from Einstein's gravity theory.  For example, we will find that
a constant energy density not only does not lead to inflation, but
it rather slows and eventually stops expansion\foot{
We shall describe the expansion/contraction in terms of the original
unrescaled ( ``$\s$ -model" ) metric. Though the
effective gravitational constant is then time dependent
, this is the metric string directly interacts with and hence the one
measured by stringy ``rods". Similar point of view was expressed in [10].
The use of this metric is appropriate at early phase of cosmological evolution
when dilaton is still massless (for a discussion of a possible
inflation in string cosmology described
 in terms of the rescaled metric  after the dilaton already
got mass see [11]). Let us note also that the use of the ``unrescaled" metric
is
particularly natural in the duality symmetric case since it has a simple
transformation law under the duality.}.
 Similarly, we will discover
 that if the
winding modes (by ``winding modes" we mean a more general notion of
 stringy states whose masses grow with a growing scale of the
universe)
are not annihilated (which may happen if they are not in thermal
equilibrium) they halt the expansion.  Using our equations we
can study some simple aspects of the cosmological scenario
proposed in [2]  in which the three dimensionality of extended space was
suggested to be a consequence of string cosmology.

The equations we consider apply to both critical and non-critical
strings and we study them both in the presence and in the absence of
matter.
The simplest solution corresponding to non-critical strings was
 first  proposed
by Myers [12] as a way to change the dimension of target
space in string theory and was then used for construction of
 cosmological solutions in [13].  These
solutions were further extended by Mueller [14]
 who found  non-trivial cosmological
 solutions for which the radii of the toroidal universe change with time.
We will show that specializing
his solution to d=2 one recovers (a region of)
the two dimensional black-hole solution, which has
been found to be related (in the leading order approximation)
 to a conformal theory [15]\foot{ As
was already pointed out in [5], the euclidean continuation of the
d=2 Myers solution coincides with the euclidean solution independently
discovered in [16].}.
We find that the more natural setup to discuss
d=2 cosmology is to go to an infinite fold cover of $SL(2,R)/U(1)$.
This will avoid getting time-like loops and
give us infinitely many universes parametrized by an integer
$n$, such that the future of particles in the universe $n$ could be in any
universe $m\geq n$.  The exact metric suggested for the d=2 Euclidean
black hole [17] which leads to a puzzle within a  black-hole interpretation
 seems perfectly consistent
once we use such a cosmological interpretation (corresponding
to the universal cover of $SL(2,R)/U(1)$).

The organization of this paper is as follows.  In section 2 we derive
the basic equations of string cosmology in the adiabatic approximation
and discuss their duality invariance property.  In section 3 we specialize
to critical strings and consider some basic elements of string cosmology
by studying the solutions of these equations.  We find a simple intuitive
way to think about the solutions based on the analogy
 with a mechanical model of a particle rolling on a potential
under the influence of a damping force.
  In section 4 we concentrate
on some aspects of non-critical string cosmology and its relation to
the d=2 black hole.  We also briefly discuss how string
one loop effects might modify the non-critical cosmological
solutions.
  In section 5 we present our conclusions. In appendix
A some aspects of the solutions to our cosmological equations are worked
out.

\newsec{General Structure of Cosmological Equations}

Let us consider
strings propagating in a large universe of spatial dimension $N$.
If $N$ is not the critical spatial dimension of strings (i.e.,
if $N< 25$ (9)  for (super)strings), we could have one of two
possibilities:  either
we have some $N_c$ of  compact internal dimensions which make up the rest
of the spatial dimensions which we take to be static and to correspond
to conformal theories on the worldsheet, or
we are dealing with a ``non-critical" string.  In other words, if we
set
 $$c={2\over 3}(d_c-N-N_c)$$
where $d_c$ is the critical spatial dimension,
 in the first case $c=0$, and in the second case $c\not= 0$.
For simplicity we take the $N-$dimensional space to be
a periodic box (torus)
 of lengths $a_i=\exp\lambda_i$ ( $i$ runs from $1$ to $N$).  However,
many of the ideas developed in this paper can easily
be extended to more general cases.

It is a well known principle in string theory that as long
as we concentrate on slowly varying fields, we can hope
to describe the global aspects of the dynamics of the theory
reasonably well by concentrating on massless modes, and keeping only
the leading derivative terms (`adiabatic' approximation).
The types of massless modes depend on  a particular choice of vacuum
backgrounds
strings are propagating in, and on whether we are considering bosonic
strings, superstrings, or heterotic strings.
  But they always include a gravitational
field and a dilaton (for simplicity we shall ignore the antisymmetric
tensor field).  If the radii are large compared to the string scale
 the leading order terms
in the low energy
expansion of the tree level gravitational-dilaton effective action of the
closed string theory are [18]
(we shall absorb the gravitational coupling constant into $\p$ and
set $\a'$=1)\foot{The volume of the internal space, if there is any,
can be gotten rid of by a shift in $\p$.}

\eqn\act{
 S_0 = - \int d^{N+1} x \sqG \  \e{-2\p}   \ [ \ c + \ R \ + 4 (D \p )^2
\ ] \ .}
Note that $\e{\p} $ plays the role of coupling `constant' in string theory.
The first question we would like to address is whether we should
trust this action at all for Planckian size spatial dimensions,
i.e., when $\lambda_i \sim 0$.  The answer is no, if we wish to
take all the modes in the above action seriously.
The above action manifestly breaks the duality symmetry of string theory.
 The reason is
that when we approach Planckian size universes the momentum modes
(which correspond to inhomogeneity in the metric and dilaton fields)
become massive, and the winding modes which wrap around the box
and were ignored previously will have to be taken into account,
as they have mass comparable to that of
 the momentum modes.  This `doubling
of the space' seems very difficult to deal with, and on top of that
we have to remember that the oscillator (higher level)
 modes are also going to become important since they will be
of the same energy as the momentum modes.

However, the following observation allows us to still
take part of the above action seriously. The point is that the
space independent parts of the above fields have zero
energy independent of the size of the universe, and thus
we can naturally concentrate on them, setting the other Planckian
modes to zero.  Note in particular that these spatially independent
modes are common to both the momentum  and winding excitations, so truncating
to these modes alone will be consistent with duality.  Therefore we should
expect that even though the above action naively seems to pick
momentum modes over winding modes, i.e., the space over the dual space,
concentrating on spatially constant
fields restores the duality symmetry.
Thus we should trust the above action as far as the overall scales
of the universe are concerned and not so much as a description
of inhomogeneities of the universe.  And in addition we should
assume that the fields are varying slowly with time.

Let us
see how the duality invariance $a_i\rightarrow 1/a_i$ is restored
in the above action once we concentrate on zero modes.  Let
us therefore consider the metric and the dilaton field of the form
$$ ds^2 = - dt^2 + \sm a_i^2 (t) dx_i^2  \  \  ,  $$
$$ a_i= {\rm e}^{\l_i(t)} \ , \ \ \p=\p(t) \ . \ \ \  $$
It is useful to
introduce the ``shifted'' (and rescaled by 2) dilaton
field $\vp$ which absorbs the space volume factor
$$ \vp \equiv 2\p - \sm \l_i \ \ ,\  \ \ \sqG\   \e{-2\p} = \e{-\vp} \   \
.$$
Then the action becomes
$$ S_0=-\int dt\  \e{-\vp} \sqrt {-\Goo}\  [ \ c - G^{00} \sm \dl_i^2
 + G^{00}\dvp^2 \ ]\ \ .$$
We have kept $G_{00}$ as we need to vary the
action with respect to it in order to get the full set of equations of
motion ( in the equations of motion $G_{ 00}$ is set  to $-1$). Dot denotes
the time derivative.  The above action is manifestly invariant
under the duality transformation [4-6]
\eqn\dual{\lambda_i \rightarrow -\lambda_i \ \ ,\ \
\p \rightarrow \p - \lambda_i \ \ , \ \  \vp \rightarrow  \vp }
(here $i$ is any number among 1,..,$N$ ; more general duality
transformations are obtained by combining these basic ones).

The above action is written in the absence of stringy ``matter".
  To discuss cosmology
it is important to incorporate matter into the system.  Assuming that the
effective string coupling is small let us take the
matter to be a gas of (free) string modes in a
 thermal equilibrium at the temperature $\beta^{-1}$.  Then
matter  contribution to the action is represented by
$$ S_m = \int dt \sqrt {- G_{00}}  \ F ( \lambda_i  , \b  \sqrt { -
G_{00} }  ) \ \ \ \ .$$
Here $F$ is the (one loop)
free energy and can be represented in terms of the one loop
string partition function in a torus of radii $\e{\lambda_i} $ and
periodic Euclidean time of perimeter $\b \sqrt{-G_{00} }$ , $\ Z=-\b F $.
The adiabaticity assumption implies that we can replace constant radii and
$\b$ by functions of time.
Taking the full action
$$S=S_0+S_m= -\int dt\   \sqrt {-\Goo}\
 [ \ \e{-\vp}  (  \ c - G^{00} \sm \dl_i^2
 + G^{00}\dvp^2 \ )  - \ F ( \lambda_i  , \b  \sqrt { -
G_{00} }  ) ] $$
 and varying it with respect
to $\lambda_i, \phi$ and $G_{00}$ and we find the following
equations\foot{Similar equations were discussed in [9] and also in [6]
where the matter was represented by the energy-momentum tensor of a
macroscopic string.  Even though we derived the above equations using
the canonical ensemble, they are valid even when the canonical
ensemble description breaks down (as is the case near Hagedorn
temperature) and we have to use the more fundamental
microcanonical ensemble. }
\eqn\econ{ c - \sm \dl_{i}^2  + \dvp^2   =  {  \evp} E  \ \ , }
\eqn\lam{ \ddl_i - \dvp \dl_i  =   \half  {\evp }  P_i \ \ ,}
\eqn\dila{ \ddvp - \sm \dl_{i}^2 =  \half  {\evp } E   \ \ ,}
where
$$E=-2{ \delta S_m\over \delta G_{00}}=F+\b {\partial F\over \partial \b}\
\ ,$$
$$P_i=-{\delta S_m\over \delta \lambda_i}=-{\partial F\over \partial
 \lambda_i}.$$
$E$ is the total energy of the matter and $P_i$ is the pressure
in the i-th direction times the volume.  These equations imply a modified
conservation law for the energy (following from the
invariance under reparametrizations of time )
$$  \  \dot E + \sm \dl_i P_i = 0 \  \ .   $$
 Since
$F=F(\l (t) , \b (t) ) \  $  this is equivalent to the
conservation of the entropy ${\cal S} = \b^2 \pa F/ {\pa \b} $,
i.e.
our matter is indeed evolving adiabatically.  This means that
for a given radius determined by $\lambda_i$ the temperature
$\b^{-1}$ adjusts itself so that ${\cal S}$ remains constant. Solving the
adiabaticity condition we can express $\b$ in terms of
$\l_i$ and hence represent $E$ as a function of $\lambda_i$ alone,
$$ E(\l) = E(\l \ , \b (\l) ) \ \ .$$
 In what follows we shall consider only this new function $E(\l ) $.
  Since entropy is not changing
we find
\eqn\pres{P_i=-{\partial E(\lambda ) \over \partial \lambda_i}\ \ .}
Note that the time derivative of \econ\ vanishes once we use
\lam\ and \dila , and so the non-trivial content of equation \econ\ is that
it
fixes  the constant of integration in terms of $c$.
The
above equations are duality invariant since $F$ is invariant
under $\lambda \rightarrow -\lambda$ for a fixed temperature ($E(\l )$  and
$\b (\l )$  are also invariant
and $P_i$ changes sign under  the duality).

Rewritten in terms of the original dilaton field $\phi$
 equations (2.3)--(2.5) take the form
\eqn\eec{ c- \sm \dlis + ( 2\dop - \sm \dl_i )^2 = \e{ 2\p } \ {\r } \  \
,}
\eqn\llam{ {\ddot \l}_j - (2\dop - \sm \dl_i ) \ \dl_j =  \half
 \e{2\p} \ { p_j}  \ \ , }
\eqn\pph{ 2\ddp - \sm  \ddl_i - \sm \dlis = \half  \e{2\p} \ {\r}  \ \ }
where $\rho =E/V$ ,  $p_i=P_i/V$  and $V= \exp {\sum
 \l_i } $ is the space volume.
To compare these equations with standard cosmological equations in which
the dilaton is ignored  consider setting $\phi$ (which is related
to the coupling constant) to be constant. Then the equation \eec\
takes the well known form (with flat space),
\eqn\stand{({\dot a\over a})^2=\dl^2={-c\over N^2-N}+G\rho}
for some constant $G$, where
$-c/(N^2-N)$ plays the role of the cosmological constant\foot{It is
amusing to note that in this setup the problem of the  vanishing cosmological
constant seems to be related to the question of being on or
off-criticality in string theory. This connection (which was already
mentioned in the past) seems
worth further study.}.  However, it is
easy to see that if $c=0$
 a solution with
$\p = \rm const \ $ is consistent with all the  three equations
only if
$$\sm p_i = \r \ \ ,$$
i.e., for a matter with the vanishing energy momentum trace, as is
effectively the case for a gas of massless particles in a thermodynamical
equilibrium. This ``radiation -type" condition is not satisfied in a high
temperature phase of string thermodynamics and is the reason for stringy
departure from Einstein's gravity.

Let us discuss some general aspects of equations \econ , \lam\ and
\dila .  It is important to notice  that the equations for
$\lambda_i$ \lam\ are the same as that of a point particle in the presence
of a time dependent potential $\ha \e{\vp} E(\lambda )$ and damped or boosted
(depending
on whether $\dvp$ is negative or positive) by a
dilaton ``friction" force.
Note that within this mechanical interpretation the ``integration constant"
$c$ in (2.3) plays the role of a fixed energy of the system.

  Also, the equation for $\vp$
can be summarized by subtracting equation \dila\ from \econ\ and defining
$y=\e{-\vp } $, which leads  to
\eqn\spi{{\ddot y}+cy={1\over 2}E\ \ .}
This equation for $y$ can be interpreted as that for the endpoint of
a spring, with spring constant $c$, and with the external force
$\ha E(\lambda )$.  These interpretations allow us to develop an
intuitive picture for the nature of the solutions to our equations.
As it turns out, the behavior of solutions will be rather different
depending on whether or not
we are describing critical strings, i.e., whether or not $c=0$.

\newsec{Critical String Cosmology}
This case corresponds to setting $c=0$.  The important qualitative
simplification that happens in this case is that since $E$ is positive
(assuming no Casimir - like negative contribution to energy) from equation
\econ\ we conclude that $\dvp$ can never become zero.  In other words,
$\dvp $ never changes sign, and thus it provides a damping {\it or} a
boosting effect in equation \lam\ for all time.   We will  consider the
damping case $\dvp <0$; the boosting case can be obtained by time
reversal.  A strong reason to consider $\dvp <0$ is that otherwise
the boosting effect of dilaton on $\lambda$ will invalidate the adiabatic
approximation used in the derivation of the above equations. Also, growing
$\vp$ together with expanding $\l$ implies the growth of the effective
coupling $\exp \p $  in contradiction with a weak coupling assumption.

{}From equation \dila\ it follows that $\ddvp>0$ and thus $\dvp$
is growing in time.  Since $\dvp$ is negative and it will never
cross zero we conclude that (as long as $E\not =0$) $\dvp$ continues
growing towards $0$ and approaches $0$ as $t\rightarrow \infty$.

For simplicity let us consider the isotropic case ,
i.e. assume that all $\lambda_i$ are equal to each other
and denoted by $\lambda$ so that $a= \exp
\l $  is  the cosmological scale.
  The equations we get can be written as
\eqn\ene{ c-N\dls +\dvps = \e{\vp} E \ \ , }
\eqn\clp{ \ddl -\dvp \dl = \half \e{\vp} P \ \ , }
\eqn\third{ \ddvp - N \dls = \half \e{\vp} E  \ \ ,}
where $P=-N^{-1} \partial E/\partial \lambda$ and for critical
strings we simply set $c=0$ in \ene .

In order to solve this system of equations we will need to specify
$E(\lambda )$  as well as provide the initial conditions for $\lambda$,
$\vp$ and $\dvp$.  $E(\lambda)$ encodes string thermodynamics by
specifying how the total energy in the box has to change as a function
of $\lambda$ in order to keep the entropy constant.
Motivated by string cosmology,
the function $E(\lambda)$ was introduced
and studied using string thermodynamics in [2] ( some properties
of $E( \l )$ were further clarified in [19],
see also [20] ). Its
basic structure is indicated in fig. 1.

This function
is duality symmetric under $\lambda \rightarrow -\lambda$.  Its structure
is well understood near $\lambda \sim 0$ and for very large $\lambda$.
In the first region near $\lambda \sim 0$,
which we call the Hagedorn region, the temperature of the strings is
very close to the Hagedorn temperature and the energy is almost independent
of $\lambda$. It is
 given by $E=T_H {\cal S}$ where $T_H$ is the Hagedorn temperature
(fixed for a given string theory) and ${\cal S}$ is the total entropy.  For
large
enough $\lambda$ temperature drops significantly compared to the Hagedorn
temperature, and the massive modes of string go out of equilibrium so that we
are  left with the massless modes.
In this  `radiation
dominated' region we thus have
the usual relation of how $E$ depends on the radius of the universe
 for a gas of massless particles.
In fact, the temperature drops as $1/a= \exp ( -\l ) $ and $E\propto T^{N+1}a^N
\propto a^{-1}$.
In other words $E=C\e{-\lambda} $ where $C$ is determined by ${\cal S}$ and the
number of massless modes.  The behaviour of $E$ in the
 intermediate region between
the Hagedorn region and the radiation dominated region is only partially
understood but it has been shown that $E$ is monotonically decreasing
(in the Hagedorn regime)
with increasing $|\lambda |$ [19], and it is natural to believe
that this is true for all $\lambda$.  So if we start with the initial
conditions
that $\dvp <0$ the decreasing of the
 potential $E$ with increasing $|\lambda |$ will force $\l$
 to slide down the potential towards increasing values of
$|\lambda |$, until we emerge at the radiation dominated era.  The
universe will continue its expansion, and in fact
it will approach the radiation dominated era of the standard cosmology.
In other words, independently of the details of $E$, or the initial
conditions, one finds that $2\p =\vp +N\l $, which determines the string
coupling constant, rapidly
 approaches a constant
and the radius ultimately grows as $a\sim A\ t^{2\over N+1}$ for
large $t$, as is the case for standard radiation dominated cosmology.
  This is discussed in appendix A.

What determines $N$, the number of extended spatial dimensions?  One
idea  to explain why $N\leq 3$ was suggested in [2].  It was
 based on the possibility that if the space expands in more than
three dimensions, the assumption of maintaining thermal equilibrium may
not be a good one, as for example the winding modes of strings will
have a hard time finding and annihilating one another. Thus they would
fall out of equilibrium and will
be around.  It was suggested there that if winding modes
are around this may stop expansion.  The universe would not expand
until it learns the lesson that
it is only possible to expand in three or smaller number of dimensions.

Let us note that Einstein's theory of gravity leaves us
uncomfortable with the above suggestion
that winding modes may prevent expansion.
Recall that cosmological expansion rate
for a flat universe with vanishing
cosmological constant satisfies $\dl^2 =G\rho$ in FRW cosmologies.
In particular any form of matter, since it contributes to $\rho$,
helps accelerate expansion.  This in particular implies
that the winding modes accelerate expansion in standard gravity theory,
rather than prevent expansion.  On the other hand the
idea in [2] was based on duality which states that what usual
matter does for the size of the universe (i.e. accelerates expansion)
the dual matter (the winding states) should do
for the dual size of the universe (i.e., accelerate contraction).
This was a puzzle which was not fully resolved in [2] . We will now see
that our equations resolve this paradox by showing that they
drastically differ from Einstein's equations in such cases
and indeed are consistent with the idea that winding modes
slow down and ultimately stop the expansion.

If the winding modes are around the energy $E$ is going to grow
with $\lambda$, as larger $\lambda$ mean bigger boxes and thus more
stretched winding strings with higher energies.   The mass of the winding
states will increase with $\lambda$ as ${\rm exp} \l  $.   $ E$ growing
with $\lambda$ means that $\ddl <0$ as $\lambda$ will
have to climb the potential $E$. So this slows the expansion and
in fact  ultimately stops it as it is shown in appendix $A$.
More generally, it is shown there that if we take for large $\lambda$
$$E\propto exp(\alpha \lambda)$$
then for $\alpha$  positive the universe reaches a maximum
size and then contracts back.  Constant matter energy density corresponds
to $\alpha =N$ and it prevents expansion even more strongly!
So we conclude that {\it dilaton suppresses the inflationary
mechanism based on constant energy density}\foot{ As was already noted in
the Introduction, this does not rule out the possibility of inflation at a
later stage of evolution when dilaton gets a mass and it is appropriate to
use the redefined metric.}.
If $\alpha$ is negative the universe
expands forever (as is the case for the radiation dominated era where
$\alpha =-1$).

The duality symmetry of $E$ implies that while the winding modes
(providing the growth of $E$ at large $\l$) oppose expansion, the momentum
modes (providing the growth of $E$ at large negative $\l$ ) oppose
contraction. As a result, in the
absence of equilibrium  the universe sees the effective $E$ represented
as a dashed curve in fig. 1 and the
radius of the Universe will be oscillating
between maximal and minimal values.

So the basic picture one is led to is a universe which oscillates in many
directions
for a while around the Planck scale (maybe within a few orders of
magnitude), until by coincidence it starts expanding in smaller
number of directions (the most likely one being the largest
possible dimension consistent with maintaining equilibrium, namely 3).
Then it expands forever and we find ourselves in the radiation dominated
era of the standard cosmology.  This may also
explain the large entropy problem.  The entropy may be large
because during this ``trial and error" period of the early universe,
as the strings were out of equilibrium we would generate ( if
this process takes long ) a lot of entropy.
In other words, the large entropy problem may be related to
having an `old' universe.

An important issue to resolve in string cosmology is to explain
the absence of massless dilatons at the present time.
One should expect that somehow
a potential is generated for the dilaton and it will have to
sit at its minimum picking up a mass.
The mechanism
of how this precisely should happen in string theory is not known yet
and is the biggest gap in connecting string cosmology to observable
cosmology.
This is presumably related to the fundamental
question of how supersymmetry breaking
takes place in string theory while maintaining vanishing cosmological constant.

\newsec{Non-critical String Cosmology}
The case of non-critical string cosmology
 is an analog of non-vanishing cosmological constant case
of cosmology in standard theory of gravity.
We shall consider the case
of $c>0$.  Cosmological solutions for the
toroidal space and non-zero $c$ in the absence of matter ( $E=0$ )
has been previously studied  in [14].
One obtains from \spi ( setting $E=0$ )
$$y=\e{-\vp} =A \ \ {\rm sin}\ 2 bt \ \ , \ \ b^2= c/4  \ \ . $$
{}From \lam\ ( with $P_i=0$) one learns that
$$ \dl_i = k_i \ {\evp } \ \ , \ \ k_i = \rm const   \ \ ,
$$
which can be easily integrated to yield
$$\l_i = \l_{i 0} +
q_i\  {\rm ln}\  {\rm tan }\ bt \ \ ,$$
subject to the condition resulting from equation \econ
$$ \sm q_{i }^2 =1 \ \ . $$
 These solutions can also be extended
to $c\leq 0$  [14] .  Note that for $c>0$
these solutions are singular, i.e., the fields blow up at finite
time.
In fact, as it is easy to see
 from \spi ,  the oscillatory
nature of solution forces $y$, which should be positive, to cross zero
in finite time ( so that $\vp$  blows up in finite time).  This
could in principle be balanced by a positive energy density appearing
on the right hand side of \spi .

For the above singular solutions the adiabatic approximation
which went into their derivation breaks down, and so we
should not trust them near the singularity.
Even so, one would expect that there should be  a
solution for any reasonable initial
condition, if we believe in completeness of string theory.
This leads us to expect that there  should be an exact
conformal theory which asymptotically reproduces Mueller's solutions.
For the case $N=1$ we will now see directly
that this is indeed the case.  In fact
the corresponding conformal theory turns out to be the
 $SL(2,R)/U(1)$ coset model which was recently linked
to the black hole geometry in [15].

For $N=1$ we have $q_1=\pm 1$. Let us take
the plus sign for $q$ as the other sign (giving the dual solution [5])
corresponds simply to shifting of time. We thus have
$$ \vp = \vp_0 - {\ln \  \rm sin} \ 2 bt   \ \ , \ \ \l =\l_0  +
{\rm ln \   tan\ } bt \ \ , \ \ \ b^2 = 4  \ . $$
This background
describes a universe with one spatial dimension with the  metric given by
(we fix a particular value for $\l_0 $ and assume that $x$ is periodic with
period $r$ , e.g. 2$\pi$ )
$$ds^2=-dt^2+\e{2\lambda} dx^2=-dt^2+ b^{-2} {\rm tan}^2bt \ dx^2.$$
The universe starts at $t=0$ with zero size and grows to infinite
size at $t=\pi/2b$.  Let us change our coordinates.  Let
\eqn\ccor{u={\rm sin }\ bt\ \e{x} \ \ , \ \ v={\rm sin}\ bt\ \e{-x} \ \ . }
We find that
$$ds^2=-b^{-2} {dudv\over 1-uv} \ \ , $$
and that the unshifted dilaton field $\p$ is given by
$$2\p =\vp +\lambda =2\p_0-2\ln \ {\rm cos}\ bt=2\p_0-{\ln } (1-uv)  \ \ . $$
This is the same metric and dilaton background found
in [15] in the region $0\leq uv\leq 1$ (here $uv={\rm sin}^2bt$), i.e.
the region between the horizon and singularity.  There is one
difference though:  here we are working on a periodic space,
so $x$ is identified with $x+r$.  This means that we have to identify
\eqn\uvt{(u,v)\sim (\alpha u, \alpha^{-1} v)}
i.e. we get a wedge in the region between horizon and
singularity (see fig. 2).  Of course the system of equation we
looked at is also valid for infinite radius $r$, which implies that
we can in fact reproduce the full region between horizon and singularity
with no identification.  In this case in order to complete the metric
we will have to add regions $I$ and $III$ as well, and we end
up getting the full black hole solution.
It is amusing that the transformation \uvt\ is an exact symmetry of
the conformal theory (in an appropriate basis it corresponds
to conjugating $SL(2,R)$ group element by the diagonal matrix
$(\alpha^{1/2},\alpha^{-1/2})$).  So we can consider an infinite
orbifold of $SL(2,R)/U(1)$ by the group generated by this symmetry
and obtain the compact universe solution.  One should note
that since $u=v=0$ is a fixed point of this transformation we end
up getting a new singularity at $u=v=0$, which in our cosmological
interpretation corresponds to the initial time $t=0$, where the
universe has zero radius.  There is another wedge branch of the solution
in regions $I$ and $III$ which has periodic time which touches
the cosmological region at the singular point $u=v=0$.

Note that in this language the duality of conformal theory
 [17, 21] corresponds (for appropriate choice of radius) to the standard
$a\rightarrow 1/a$ duality of our one dimensional universe,
i.e.  $bt\rightarrow {\pi \over 2} -bt$ changes $uv\rightarrow 1-uv$
and $\l \rightarrow - \l $.

Another aspect of Mueller's solution is that we can continue time
past the singularity at $t=\pi/2b$.  In particular, if we introduce
$$ u_1=-{\rm sin}\ bt \ \e{x} \ \ , \   \   v_1=-{\rm sin}\ bt \  \e{-x} $$
we get an identical copy of the interior region of black hole
(on the overlap $u_1=-u \ , \ \ v_1=-v$)
where the universe now starts at the lower singularity, and evolves
till $t=\pi/b $ at which it reaches zero size, it grows
as $t$ increases until it becomes infinite in size again at $t=3\pi/2b$.
Then again we use the coordinate transformation
$$u_2={\rm sin} \ bt \ \e{x} \ \ , \  \ v_2={\rm sin} \ bt \ \e{-x} $$
where the
radius shrinks to zero size at $t=2\pi/b$.  By now the universe
has undergone two oscillations, but we have covered $SL(2,R)$ only
once. For example, if we take $x=0$, the corresponding element
of $SL(2,R)$ , as we evolve in time, is represented by the matrix
$$\left(\matrix{{\rm cos}\ bt&{\rm sin}\ bt\cr -{\rm sin}\ bt&{\rm cos}\ bt
\cr}\right)$$
(it is known that the $SL(2,R)/U(1)$ has two copies of the interior region
of black hole--to get only one we could have used $PSL(2,R)/U(1)$). But now
we will {\it not} identify $(u_2,v_2)$ with $(u,v)$, as that would
have given us periodic time.  Instead we continue forever.  This means
that in the above matrix realization of $SL(2,R)$ we are
not identifying $t \rightarrow t+2\pi/b$.  In other words
the cosmological interpretation suggests that we go to an infinite fold
cover of $SL(2,R)$.  In this way we end up with a universe which
had no beginning and no end, and it undergoes infinitely many
oscillations.  Again in this picture we have the option of choosing a
finite radius or infinite radius for our space.  If we
have a finite radius, then we will have to restrict to the wedge
region between the horizon and singularity
 shown in fig.2, otherwise we will have to add the regions
$I$ and $III$ to complete the metric.  In this latter interpretation
an observer in region $I$ or $III$ of one of these universes, may decide
to enter the next universe.  The Penrose diagram for this series
of universes is shown in fig.3.  It is amusing to note that in
the Euclidean version of each of these universes we have a periodic
coordinate which suggests  a thermal bath interpretation.
This suggests, even in the cosmological picture, we get radiation
from the horizons.  It would be interesting to study this further.

Now we ask the question if this one dimensional cosmological
solution should be taken seriously.  Naively the answer would be no,
because we have violated the assumption of adiabaticity of the fields
near the singularity, when the size of the universe is infinitely
large.  However, since we know there is an exact conformal theory
with the correct asymptotic behavior, we are led to expect that there
exist a corrected  metric which gives the exact answer.  Based
on comparison with the coset model an exact metric for the Euclidean black
hole was suggested in [17]. As was checked in [22] the corresponding
metric - dilaton background solves the $\s$ model conformal invariance
conditions in the 3-loop approximation.
When one analytically continues the conjectured metric
to the region between horizon and singularity it reads
$$ds^2=-dt^2+ b^{-2}{{\rm tan}^2bt\over {1+q{\rm tan}^2bt}}dx^2 \ \ ,
\ \ \ q=2/k=8/9 \ \ .  $$
  If this metric is indeed exact, it suggests
that the space-time is actually nonsingular at $t=\pi/2b$  so that
we see no singularity in the metric! In the finite radius
scenario (where we introduce an additional singularity at $u=v=0$)
we have a universe starting at zero radius, reaching a finite
maximum radius at finite time and then contracting back to zero radius
and continuing this indefinitely.
The conjectured form
for dilaton $\vp$ is unchanged from the leading approximation and
is thus still singular at $t=\pi/2b$.  The cosmological interpretation
of the black hole gets rid of a puzzling consequence of this conjectured
exact metric:  if analytically continued to the region behind black hole
singularity
 the metric becomes {\it Euclidean} between $1\leq uv \leq   9$
with a singularity at $uv=9$.  This is avoided in our interpretation
because the region beyond the singularities are never reached (they are
connected to universes labeled by $\pm \infty$).

It would be interesting to find exact conformal field theory
generalisations of higher dimensional  Mueller--type cosmological
solutions\foot{
In such a case it would be natural to allow the full moduli of toroidal
compactifications to become involved.  The picture is a
more or less straight-forward generalization of the
equations we have considered, and we would simply have to study
geodesic equations on the corresponding moduli spaces of
toroidal compactifications.}.
The coset theories like $SO(d,1)/SO(d-1,1)$ which should be analogs of
the (anti) DeSitter backgrounds (see Bars [21] and refs.
there) does not, however, correspond to a flat $N$-space. The
generalization of the
leading order conformal invariance equations to the case of
isotropic homogeneous metric with a curved space (with positive, zero or
negative curvature $k$ ) is given by (cf. (3.1)--(3.3) )
$$ c -N\dls +\dvps = -k N(N-1) \e{-2\l}\ , $$
$$ \ddl - \dvp \dl = - k (N-1) \e{-2\l}\ , $$
$$ \ddvp - N\dls = 0 $$
(the absence of the correction in the third equation is due to the fact
that the ``potential" in the present case is ``classical", i.e. does not
depend on the dilaton).Let us assume that $c<0, k>0$.
 Then at large negative $\l$ (small times) the solution is approximately
 given by the  DeSitter metric and a constant dilaton $\p=\ha (\vp +N\l)$.
In this limit, however,
the adiabatic approximation is not fully reliable.
At large positive $\l$  the effect of the space curvature becomes
irrelevant and the solution approaches  the Mueller's isotropic solution
$$ \l= \l_0 + N^{-\ha} {\ln \tanh }\ bt  \rightarrow \const \ \ ,
\vp \rightarrow  - bt  \ \ . $$
So the universe is born at $t=0$ as a $d=N+1$ DeSitter space
and evolves into the product of a  time line and an $N$--sphere at large $t$.
This appears to be a direct higher dimensional (Minkowski signature)
analog of the $d=2$  (euclidean) ``cigar" metric.

So far we have discussed non-critical string cosmology in absence
of matter.  It is natural to ask how can one introduce
`thermal' non-critical matter, in order to have a richer non-critical
string cosmology.  It turns out that this can be done at least
formally.  The usual way we consider thermal ensemble in field theory
is by making the time Euclidean with period $\beta$.  For non-critical
strings the Liouville field plays the role of time and this suggests
making it Euclidean and periodic with period $\beta$.   The computation
of thermal free energy, which is a one loop computation generalizing
that of [23] can be carried out
explicitly for the non-critical strings $c_m\leq 1$.  For $c_m=1$
noncritical strings at radius $a$, for
example, one finds that the necessary modular integral has been evaluated
in [24] in a completely different context. This is related
to the $S^1 \times
S^1 $ partition function integrated over the fundamental domain and
turns out to be the same as that for
the critical $N=2$ strings partition function
[25] ( or, equivalently, for
$c_m=1$ non-critical $N=2$ string partition function).
One finds for the free energy
$$\beta F ={\rm ln}[ \ \tau_2 \eta^2(\tau ){\bar \eta}^2(\bar \tau)\ ]
+(\tau
\rightarrow \rho )$$
where $\tau = i \tau_2 =i \beta a/8\pi $, $\rho =i \beta /8\pi a
$ and $\eta$ is the Dedekind
eta function.  Solving the adiabaticity condition
 one can obtain the energy $E(a)$ from
the above expression for $F$.  For large entropy and small $a$ we have
$$E(a)={A\over a+ a^{-1}}$$
where $A$ depends on the entropy.  The profile for $E(a)$
is very similar to that of critical strings (fig.1) except
that there is no flat Hagedorn region and we only get
a maximum at $\lambda =0$.  This is to be expected
since at $c_m=1$ there is no exponential degeneracy of states, as
there is only one massless scalar (and some additional discrete
states [26]) and so there is no limiting temperature.  In fact
it has been suggested  [27]
that introduction of temperature for non-critical strings
might lead to a more clear phase diagram picture for non-critical
strings.
Note that for large $a$,  $\ E(a)$
behaves like the energy for that of a massless particle ($E\propto
1/a$).  For yet larger radii, the temperature drops to zero, but
$E$ does not go to zero.  It becomes negative and
behaves as
$$E=-{1\over 24 }(a+{1\over a})  \ \ . $$
This is a Casimir - type contribution which is independent of the temperature.
It is in accord with the result of [23]
which corresponds to computing the free energy at zero temperature.  Note that
this Casimir effect is absent in critical (super)-strings, as the
cosmological constant vanishes at one loop.  Putting
the potential
 $E(a)$ on the right hand side of our equations will modify Muellers
solution.  The intuition based on the $\lambda$ rolling on the
potential hill $E(a)$ implies that the solutions are
still singular (and blow up in finite time).

These ideas can be applied to $c_m<1$ as well.  In this case we
have a one dimensional field theory (corresponding
to the Liouville  field), which we
can identify with a space coordinate $x$. The relevant equation to solve for
the dilaton at the  tree level is
$\ c-\vp'^2=0 \ $ which
 gives rise to the familiar linear dependence of the  dilaton:
  $\vp= -{\sqrt c}\  x$.  However, the one loop partition function
[23]
modifies the above equation to
$$c-\vp'^2=- h \evp$$
where $h$ is a positive constant depending on which minimal model
we are dealing with.
Solving the above equation we find
$$\vp=\vp_0-{\rm ln\ sinh}^2{{\sqrt c}\ x\over 2} \ \ . $$
This suggests a truncation of the Liouville space to a  half-line.  Again
we are going beyond the validity of adiabatic approximations
but we are getting the hint that the linear dependence
of dilaton is modified at one loop.

\newsec{Conclusions}
We have discussed some aspects of string cosmology
taking into account the dynamics of the dilaton field
which is needed to restore the duality symmetry in string theory.
We have considered both  critical and  non-critical
strings, the non-critical strings being the analog of having
a non-vanishing cosmological constant in the Einstein theory.

For critical strings we find a major modification of the description of the
 early universe.
In particular, winding string states, if not annihilated,
 can halt the expansion of the universe.  This is against the
intuition based on standard gravitational theory where a
constant energy density leads to inflation.  We have seen that
string dynamics is consistent with the following picture of cosmology
which needs to be verified.
A nine dimensional spatial universe of Planck size which we take to be a
periodic ``box" ,
at the Hagedorn temperature, expands in all directions.  Not being able
to get rid of winding modes the expansion stops and the universe
contracts to even smaller than the Planck scale. Now the momentum modes
cannot annihilate one another and the contraction stops. The universe
starts to grow once again.  The universe
oscillates for a while in this fashion.  In this way one can in principle
generate a lot of entropy.
Fluctuations cause different
directions to expand at different times. One of these fluctuations
leads to three dimensions expanding, and now the winding modes can
get annihilated and the universe expands, getting rid of the winding modes.
As the temperature drops, the
 Planckian string states  get suppressed
by the Boltzman factor and we are left with the massless modes
in a three dimensional expanding universe, which looks very much
like the radiation dominated era of the standard cosmology.  It is
remarkable that the radiation dominated era of the
 standard cosmology emerges with no fine tuning.

How can one test this picture?  One idea is to treat strings classically
as one does for  cosmic strings.
So we can study the evolution of a universe filled with these strings
which interact with each other by cutting and rejoining. The main
difference with the cosmic strings is that we have a different evolution
equation for the universe, and the coupling constant for the
cutting and rejoining which is related to dilaton field is also evolving.
This approach  will presumably answer questions of entropy production
and homogeneity of the early universe in the cosmological scenario
described above.

We have also studied non-critical string cosmologies, which seem
qualitatively rather different from the critical string case.
In particular we have related the empty universe cosmology in $d=2$
to the two dimensional
 `black hole' solution (more precisely to its universal cover).
Could it be that stringy black-holes in 4d behave in a similar fashion?
In particular, as we cross the horizon we enter another universe
instead of encountering the singularity? It would be remarkable if the
``cosmology--black hole" connection discovered in two dimensions has higher
dimensional analogs.  Indeed, the fact that the singularity
of standard 4d black hole solution is also spacelike, i.e. appears
at a given time rather than space, suggests a cosmological interpretation.

To discuss non-critical string cosmologies with matter we have
to introduce the notion of temperature for non-critical strings.
We have made a suggestion of how this may be done (by a formal wick rotation
of Liouville field and making it periodic with period $\beta$ as one
does for $c_m=25$ in order to define critical string thermodynamics
)\foot{At
higher genus due
to the coupling of Liouville field to the curvature,
this may be consistent only if $\beta$ is quantized.}.
  Using this definition
we have computed the thermal properties of $d=1$ matter
(which can be easily generalized to $d<1$ as in [23]).  It is an
interesting question to see
if there exists a matrix model analog of this finite temperature
interpretation (for example are unitary matrix models related to
thermalizations
of standard  matrix theories?).

We would like to thank M. Bershadsky, A. Guth, S. Jain, J. Polchinski,
M. Tsypin, G. Veneziano
 and E. Verlinde for interesting discussions. A.T. would like to
acknowledge J.Bagger for hospitality at Johns Hopkins University
and various kinds of help.  The research of
A.T. was supported by NSF grant PHY-90-96198 and that of C.V. was
supported in part by Packard Foundation and NSF grants PHY-89-57162
and PHY-87-14654.
\appendix{A}{}
In this appendix we discuss the solution to the
equations \ene ,\clp ,\third\ with $c=0$ and
\eqn\boe{E\propto exp(\alpha \lambda)}
starting with initial conditions $\dvp <0$, $\dl >0$.  In particular
we show that if the total energy grows with volume, i.e., if $\alpha >0$
the universe stops expanding in finite time and starts contracting,
i.e., we reach $\dl =0$ in finite time as $\lambda $ is decelerating.
If $\alpha <0$ the universe expands forever and if $\alpha =0$ the
universe comes to a halt, without turning back.
We will show in addition that
if $\alpha =-1$, as would be the case for radiation dominated era,
the radius eventually grows as in standard Einstein's gravity
in the radiation dominated era.

One can reach the  above conclusions more or
less immediately by using the ideas developed in section $3$.
As discussed there starting with $\dvp <0$, we conclude that $\dvp$
cannot become zero, and thus it will always remain negative.  Since
$\ddvp >0$ from \third\ we conclude that $\dvp $ approaches zero
and gets there at $t=\infty$.
Equation \clp\ implies that we can treat
$\lambda$ as the position of a particle rolling
on a potential described by $\e{\vp } E(\lambda )/2N$  with a damping
term (because of $\dvp$ terms in \clp ).
The potential is getting weaker as its overall size is modulated
by $\evp$ which is decreasing because $\dvp <0$ and the
damping term gets weaker as $\dvp$ is approaching zero.
For $\alpha <0$ it is clear that since $E$ is decreasing
with $\lambda$, $\lambda$ will continue to increase indefinitely.
For $\alpha >0$, since the energy increases with $\lambda$, the
universe cannot grow indefinitely, and will reach a maximum and start
contracting.  This is almost obvious, were it not for the fact
that the energy is modulated with $\evp $ which is
getting smaller.  So the universe could conceivably continue expanding
at ever slower rates.  This we will show is not the case. We shall
find by a careful analysis of these equations that the universe
reaches a maximum size and starts contracting after that.
  The fact that
for $\alpha= 0$ (i.e., constant energy), the universe
reaches a maximum size and stops is also clear from
\clp\ as the potential is flat in this case, and the damping
term will eventually stop $\lambda$.

First we note that $P=-N^{-1} \partial E/\partial \lambda =
-\alpha E/N$.  Then we can eliminate ${\evp} E$ from equations
\ene ,\clp , \third\ and obtain
$$\ddl-\dvp \dl ={-\alpha \over 2N} (-N\dl^2+\dvp ^2)\ \ ,$$
$$\ddvp -N\dl^2={1\over 2}(-N \dl^2+\dvp^2)\ \ . $$
Now we can represent these equations in the  first order form
by defining
$$l=\dl \ \ ,\ \ \ f=\dvp$$
to obtain
$${\dot l}={\alpha l^2\over 2}+lf -{\alpha f^2\over 2N}\ \ , $$
\eqn\linee{{\dot f}={Nl^2\over 2}+{f^2\over 2} \ \ . }
The initial conditions are $f<0$ and $l>0$.  Actually we
have another restriction, which comes from positivity of $E$ in \ene\

$$|f|\geq |{\sqrt N} l|\ \ .$$
The important point to note in solving \linee\ is that these
equations are homogeneous. In other words, if we rescale
$(l,f)\rightarrow r(l,f)$ and rescale $t\rightarrow r^{-1}t$
we get a new solution.  This  means, in particular, that
if we look for solutions with
\eqn\slo{{df\over dl}={f\over l}}
we get straight lines in the $(f,l)$ plane
which pass through $(0,0)$. Studying these particular
solutions will give us a handle to study
a  qualitative behavior of all solutions.
These straight line solutions can be easily found by
noting that ${df/dl}={\dot f}/{\dot l}$. Solving \slo\ we find
three solutions
$${f\over l}=\pm {\sqrt N}\ ,\ \ {N \over \alpha} \ \ . $$
By the rescaling argument, $f$ and $l$ can approach zero
only along the above lines.  So in the region $\alpha >0$,
if we start with the initial conditions $l>0 \ , \ f<0\ , \
|f|>{\sqrt N}l\ $,
the only possibility for not crossing  the $l=0$ line is for the flow
to approach zero through the line $f/l=-{\sqrt N}$.  But a simple check
of the equation shows that this line is repulsive, i.e., the flows are
driven away from it.  They will be attracted to the next fixed line which
is in the $l<0$ region.  They will cross the $l=0$ line which
means that at finite time $\dl =0$.  According to \clp\  $\ddl$
is not zero so  the universe bounces back
and $\lambda$ starts to decrease.

For $\alpha =0$, the solution with $l=0$ is an attractive
solution and all solutions will approach it.  This simply
means that if we start with $\dl >0$  the expansion will eventually
stop because of the damping term.  In fact, it is easy to write
the general solution in this case: using \spi\ and \clp  ( with $P=0$ ) we
get
$$\e{-\vp} ={E_0t^2\over 4}-{NA^2\over E_0} \ \ ,
\ \l =\l_0+{1\over {\sqrt N}}{\rm ln}{t-2{\sqrt{N}}A/E_0\over
t+2\sqrt{N} A/E_0}\ \ .$$
This solution approximately describes the evolution
in the Hagedorn region (in fact, $E$ is not
quite constant and decreases with increasing $\lambda$ so the
expansion never quite stops).

If $-{\sqrt N}<\alpha <0$, the line $f/l=N/\alpha$
is an attractive solution and all solutions will approach it.  Note that
for $\alpha=-1$ which corresponds to
 the case of the radiation dominated era, the
fixed line is $f/l=-N$. For the fixed line
solution $2\dot{\p }=\dvp+N \dl=f+N \ l=0$, i.e. the original dilaton
$\p$ is constant and we get the standard (flat) FRW - type cosmology
in the radiation dominated era,  $a\sim t^{2/(N+1)}$.  An important
point is that this is an attractive solution, i.e. all solutions
(with initial $\dot \vp <0$) will eventually approach it !  So at late times
we get the standard radiation dominated era of cosmology without any fine
tuning.
If $\alpha <-{\sqrt N}$ the attractive solution will be
$f/l =-{\sqrt N}$.

\vfill\eject

\centerline{\bf References}

{\settabs 16\columns
\+ 1. & K. Kikkawa and M. Yamasaki,
      Phys. Lett. B149 (1984) 357; \cr
\+    & N. Sakai and I. Senda,  Progr. Theor. Phys. 75(1986)692;\cr
\+    & V. Nair, A. Shapere, A. Strominger and F. Wilczek,
   Nucl. Phys. B287(1987)402;\cr
\+    & B. Sathiapalan, Phys. Rev. Lett. 58 (1987) 1597. \cr
\+ 2. & R. Brandenberger and C. Vafa, \NP B316(1988) 391. \cr
\+ 3. & A.R. Bogojevic, Brown Univ. preprint 1988, BROWN-HET-691;\cr
\+    & E. Witten, \PRL 61(1988)670 ;\cr
\+    & A.A. Tseytlin, \NP B350(1991)395. \cr
\+ 4. & P. Ginsparg and C. Vafa, \NP B289(1987)414;\cr
\+    & T.H. Buscher, \PL B194(1987)59 ; \PL B201(1988)466; \cr
\+    & T. Banks, M. Dine, H. Dijkstra and W. Fischler, \PL B212(1988) 45;\cr
\+    & G. Horowitz and A.A. Steif, \PL B250(1990)49; \cr
\+    & E. Smith and J. Polchinski, \PL B263(1991)59.  \cr
\+ 5. & A.A. Tseytlin, \MPL A6(1991)1721. \cr
\+ 6. & G. Veneziano, preprint CERN-TH-6077/91. \cr
\+ 7. & A.A. Tseytlin, ``Space--time duality, dilaton and string
cosmology'', \cr
\+    & Proc. of the First International A.D. Sakharov
Conference on Physics, Moscow \cr
\+    &  27- 30 May 1991, ed.  L.V. Keldysh
et al., Nova Science Publ., Commack,\cr
\+    &N.Y. , 1991. \cr
\+ 8. & H. Nishimura and M. Tabuse, \MPL A2(1987)299; \cr
\+    & J. Kripfganz and H. Perlt, Class. Quant. Grav. 5(1988)453; \cr
\+    & N. Matsuo, Z. Phys. C36(1987)289; \cr
\+    & Y. Leblanc, \PR D10(1988)3087; \cr
\+ 9. & M. Hellmund and J. Kripfganz, \PL B241(1990)211. \cr
\+ 10.& N. Sanchez and G. Veneziano, \NP B333(1990)253; \cr
\+    & B.A. Campbell, A. Linde and K.A. Olive, \NP B355(1991)146.  \cr
\+ 11.& M.C. Bento, O. Bertolami and P.M. Sa, \PL B262(1991)11 .\cr
\+ 12.& R. Myers, \PL B199(1987)371 .\cr
\+ 13.& I. Antoniadis, C. Bachas, J. Ellis, D. Nanopoulos,
\PL B211(1988)393;\cr
\+    & \NP B328(1989)115. \cr
\+ 14.& M. Mueller, \NP B337(1990)37. \cr
\+ 15.& E. Witten, \PR D44(1991)314.  \cr
\+ 16.& S. Elitzur, A. Forge and E. Rabinovici, \NP B359(1991)581; \cr
\+    & G.Mandal, A.M. Sengupta and S.R. Wadia, \MPL A6(1991)1685. \cr
\+ 17.& R. Dijgraaf, H. Verlinde and E. Verlinde, Princeton
 preprint PUPT-1252/91. \cr
\+ 18.& J. Scherk and J.H. Schwarz, Nucl. Phys. B81(1974)118; \cr
\+    & E.S. Fradkin and A.A. Tseytlin, \NP B261(1985)1; \cr
\+    & C.G. Callan, D. Friedan, E. Martinec, M.J. Perry, \NP
B262(1985)593. \cr
\+ 19.& N. Deo, S. Jain and C.-I. Tan, \PR D40(1989)2626 ;\cr
\+    & \PL B220(1989)125; Harvard preprint  HUTP-91/A025. \cr
\+ 20.& M. Bowick and S. Giddings, \NP B325(1989)631 .\cr
\+ 21.& A. Giveon, LBL preprint LBL-30671/1991; \cr
\+    & E. Kiritsis, LBL preprint LBL-30747/1991; \cr
\+    & I. Bars, U. of Southern California preprint USC-91/HEP-B3 (1991).\cr
\+ 22.& A.A. Tseytlin, \PL B268(1991)175. \cr
\+ 23.& D.J. Gross and I.R. Klebanov, \NP B344(1990)475; \cr
\+    & M. Bershadsky and I.R. Klebanov, \PRL 65(1990)3088; \cr
\+    & N. Sakai and Y. Tanii, Int. J. Mod. Phys. A6(1991)2743.   \cr
\+ 24.& L. Dixon, V. Kaplunovsky and J. Louis, \NP B355(1991)649.\cr
\+ 25.& H. Ooguri and C. Vafa, \NP B361(1991)469.\cr
\+ 26.& D.J. Gross, I.R. Klebanov and M.J. Newman, \NP B350(1991)621; \cr
\+    & A.M. Polyakov, \MPL A6(1991)635. \cr
\+ 27.& C. Vafa, Int. Jour. of Mod. Phys. A6(1991)2829.\cr

\vfill
\nfig\name{The solid curve represents the adiabatic variation of energy
$E$ as a function of scale $\lambda$.  Note the duality symmetry
$\lambda \rightarrow -\lambda$.  Near the Planck scale
($\lambda \sim 0$) the energy is more or less independent of scale,
as we are in the Hagedorn regime of string thermodynamics.  For
large radii, we enter the radiation dominated era and $E\propto{\rm exp}
-\lambda$. The dashed curve represents the effective energy if
the winding modes (and momentum modes for $\lambda <<0$) fail to
annihilate.}
\nfig\nnm{The 2d black hole geometry is represented here.
The cosmological solution correspond to regions $II$ and $IV$.
If we wish to deal with a compact universe, we
are limited to the wedge drawn here.  In this case
the curve drawn in region $II$ inside the wedge
represents the space, with its endpoints identified.}
\nfig\nmn{The `accordion-like' Penrose diagram of 2d cosmology.
An observer in universe $n$ can end up in any universe $m\geq n$.
  The regions inside the curves represent the universe if
we wish to consider compact space.}

\listfigs

\bye

\end